\documentclass[jkps,preprint,twocolumn,fleqn,showpacs,showkeys]{revtex4}
\usepackage[pdftex]{graphicx}
\usepackage{amssymb}
\usepackage{amsmath}
\usepackage{bm}
\begin{document}
\title[]{Superheavy Dark Matter in Light of Dark Radiation}
\author{Jong-Chul \surname{Park}}
\email{log1079@gmail.com}
\author{Seong Chan \surname{Park}}
\email{s.park@skku.edu}
\affiliation{Department of Physics, Sungkyunkwan University, Suwon
440-746, Korea}

\begin{abstract}
Superheavy dark matter can satisfy the observed dark matter abundance if the stability condition is fulfilled. 
Here, we propose a new Abelian gauge symmetry ${\rm U(1)}_H$ for the stability of superheavy dark matter as the electromagnetic gauge symmetry to the electron.
The new gauge boson associated with ${\rm U(1)}_H$ contributes to the effective number of relativistic degrees of freedom in the universe as dark radiation, which has been recently measured by several experiments, {\it e.g.}, PLANCK. We calculate the contribution to dark radiation from the decay of a scalar particle via the superheavy dark matter in the loop. Interestingly enough, this scenario will be probed by a future LHC run in the invisible decay signatures of the Higgs boson.
\end{abstract}

\pacs{11.30.-j, 12.90.+b,95.35.+d,98.80.Cq }

\keywords{Superheavy dark matter, Dark radiation, Higgs invisible decay, Planck, LHC}

\maketitle

\section{Introduction}

Although dark matter (DM) is one of the most important constituent of the universe~\cite{DM-Review}, its nature is still almost unknown. 
Indeed, we still have no concrete evidence for the weakly interacting massive particle (WIMP) from any experiments including the Large Hadron Collider (LHC) and DM direct searches below $\mathcal{O}$(TeV) even though the WIMP or ``heavy neutrino''~\cite{Lee:1977ua} has been regarded as a promising candidate since the late 1970s.
On the other hand, heavier DM particles, $\gg \mathcal{O}$(TeV), have been relatively less studied due to a mass limit from the so-called {\it unitarity bound}. 
If this bound is applied on the DM relic density, for a Majorana fermion, one can obtain the following relation~\cite{Griest:1989wd}:
\begin{equation}
\Omega_{\rm DM}h^2 \geq 1.7 \times 10^{-6} \sqrt{x_f} \left[\frac{m_{\rm DM}}{{\rm 1 TeV}} \right]^2,
\end{equation}
where $x_f = m_{\rm DM}/T_f$ and $T_f$ is the freeze-out temperature. The result is similar for a scalar, while $\Omega_{\rm DM}h^2$ is a factor of 2 larger for a Dirac fermion.
To avoid the overclosure of the universe, we should set $\Omega_{\rm DM}h^2 \leq 1$. Thus, we find the upper limit on the DM mass, $m_{\rm DM} \lesssim 340 {\rm TeV}$ for $x_f\approx 28$. 
If the current bound on the DM abundance is used, $\Omega_{\rm DM}h^2 \approx 0.12$~\cite{Planck2013}, the constraint can be stronger, $m_{\rm DM} \lesssim 120 {\rm TeV}$.
In addition, other crucial challenges exist for heavy DM particles:
\begin{itemize}
\item  {\it Instability of DM:} A heavier particle is generally more unstable if no symmetry principle precludes its decay.
\item {\it Lack of testability:} In currently on-going or future experiments, heavier DM well above $\mathcal{O}$(TeV) is difficult to test~\cite{DM-Review}.
\end{itemize}

One can evade the unitarity bound in the case of superheavy DM with mass $m_{\rm DM} \approx 10^{12-14}$ GeV, which is generically called as {\it WIMPZILLA}.
Independently of the nongravitational interactions, WIMPZILLA particles can be produced in various ways and satisfy the observed DM abundance: {\it i.e.}, from the expansion of the background spacetime at the end of the inflation or in the bubble collision in a first-order phase transition.
For more details, see, {\it e.g.}, Refs.~\cite{WIMPZILLA} and \cite{WIMPZILLA2}.

On the other hand, for WIMPZILLA, the instability problem gets worse due to the extremely high mass. 
In the presence of gravity, a global symmetry does not work, so all the stable particles need gauge symmetries~\cite{Kallosh:1995hi, Banks:2010zn}. 
In this work, for the stability of WIMPZILLA, we introduce an Abelian group U(1)$_H$, which similarly acts as U(1)$_{em}$ for the electron in the standard model (SM).
Moreover, a new massless gauge boson or hidden photon associated with U(1)$_H$ could provide a possibility to test WIMPZILLA. 
The new boson does not directly interact with the SM sector, thus, we regard this new boson as dark radiation. 
Dark radiation could have left evidence of its presence in various stages of cosmological history. 
Recently, observations show that additional relativistic particles may have existed at the Big Bang nucleosynthesis (BBN) time and at the recombination era shown in cosmic microwave background radiation (CMBR).
In the rest of this work, we will introduce a model with WIMPZILLA and study the testability of this WIMPZILLA via the associated dark radiation in BBN and CMBR~\cite{Menestrina:2011mz, Fischler:2010xz} and the possible signatures at the LHC~\cite{Kim:2009qc}.

\section{The Model}

We propose the minimal model including two hidden sector particles: a Dirac fermion $\psi$ and a scalar $\phi$. 
The fermion $\psi$, the DM candidate, is only charged under the `hidden' Abelian gauge symmetry U(1)$_H$. 
The scalar particle $\phi$ is neutral under U(1)$_H$ as well as the SM gauge symmetries, but the late time decay of $\phi$ into hidden photons is responsible for dark radiation. 
The gauge invariant Lagrangian\footnote{The kinetic mixing term ($\sim F_{\mu\nu}F_H^{\mu\nu}$) can be another source of interactions between the hidden and the SM sectors~\cite{KineticMixing, KineticMixing2} if there exists a bi-charged particle under the gauge symmetries of both sectors. 
However, such a particle is absent in this model.} is given by
\begin{eqnarray}\label{Interactions}
&{\cal L}& \supset {\cal L}_{SM}+{\cal L}_\phi
+ \lambda_{\phi H} \phi^2 (H^\dagger H)
- \frac{1}{4}\, F_{H \mu\nu}F_H^{\mu\nu}~~\nonumber\\
&-& y_\psi \phi \overline{\psi}\psi + i
\overline{\psi}\gamma^\mu (\partial_\mu - i g_H A^H_\mu) \psi
- m_\psi \overline{\psi}\psi,~~
\end{eqnarray}
where ${\cal L}_\phi = \frac{1}{2}[(\partial \phi)^2 -m_\phi^2 \phi^2]$ and the `Higgs portal' interaction~\cite{HiggsPortal} is allowed.
The new fermion can be naturally heavy enough due to the vectorlike mass term and, thus, is consistent with the WIMPZILLA picture. 
For the late time decay, however the new scalar is required to be much lighter than the WIMPZILLA, which provides a similar hierarchy problem as the SM Higgs boson.  
If a (super)symmetry is allowed in the hidden sector, the scalar boson can be protected from the radiative correction. 
In addition, unnecessary terms such as $\phi H^\dagger H$ are automatically forbidden. 
The details will be given in Ref.~\cite{future}.

\begin{figure}
\includegraphics[width=7.0cm]{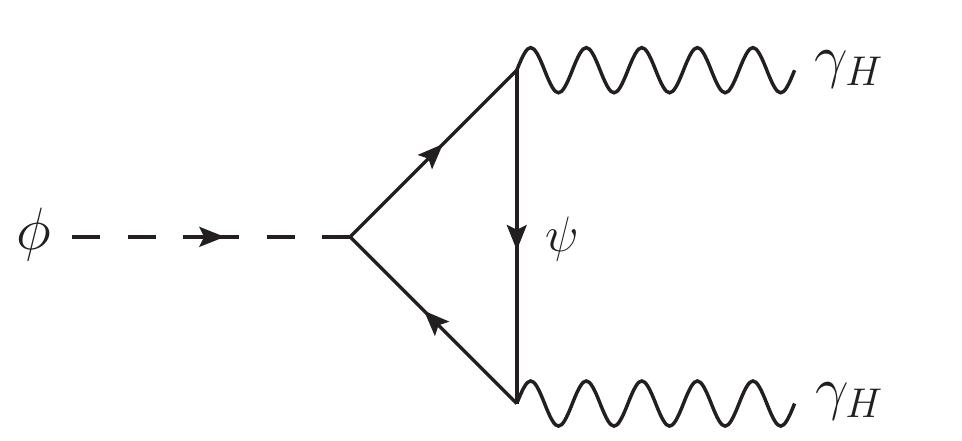}
\caption{Scalar particle $\phi$ decaying into two hidden photons through the WIMPZILLA $\psi$ loop.}
\label{fig1}
\end{figure}

The scalar particle decays into two hidden photons through a diagram with the virtual $\psi$ in the loop, as shown in Fig.~\ref{fig1}. 
The decay rate of $\phi$ is strongly suppressed by the superheavy mass of WIMPZILLA $\psi$:
\begin{eqnarray}
\frac{1}{\tau_{\phi \to 2\gamma_H}} = \Gamma_{\phi \to 2\gamma_H}\approx \frac{(\alpha_H y_\psi)^2}{144 \pi^3} \frac{m_\phi^3}{m_\psi^2}\,,
\end{eqnarray}
where $\alpha_H = g_H^2/4\pi$. 
With appropriate parameters, the life time of $\phi$ can lie in an interesting epoch: the BBN epoch, $t_{\rm BBN} \approx \mathcal{O}(0.1-1000){\rm s}$, or the CMB epoch, $t_{\rm CMB} \approx 1.2 \times 10^{13} {\rm s}$.
Consequently, we could test this WIMPZILLA model by observing dark radiation in these epochs.

\section{Observational Limits}

Now, we will consider the contributions to dark radiation observed in BBN and CMB data and the invisible decay of the Higgs at the LHC.

\subsection{Dark Radiation}

The recent observational results on the effective number of relativistic degrees of freedom, $N_{\rm eff}$, are as follows:
\begin{eqnarray}
N^{\rm CMB}_{\rm eff} &=& 3.30 \pm 0.27 \quad \text{(Planck 2013~\cite{Planck2013})},~ \nonumber \\
N^{\rm CMB}_{\rm eff} &=& 3.84 \pm 0.40 \quad \text{(WMAP9~\cite{WMAP9})},~~ \label{eq:neff}
 \\
N^{\rm BBN}_{\rm eff} &=& 3.71^{+0.47}_{-0.45}  \quad \quad \text{(BBN~\cite{BBN})}, \nonumber
\end{eqnarray}
which show some deviations, in particular, in WMAP9 and the BBN results compared to the SM expectation, $N^{\rm SM}_{\rm eff} = 3.046$~\cite{NeffSM}. 
One should note that adding the independent $H_0$ measurement to the Planck CMB data provides $N_{\rm eff}=3.62\pm 0.25$, which corresponds to a $2.3 \sigma$ deviation from the SM expectation~\cite{Planck2013}. 
Indeed, a new relativistic degree of freedom, the so-called {\it Dark radiation}, alleviates the tension between the CMB data and $H_0$.

In this model, the late time decay of $\phi$, in contact with the SM sector through the Higgs portal interaction, provides sizable dark radiation contributions at the BBN and the CMB epochs. 
If $\phi$ has never dominated the expansion of the universe, the extra contribution to $N_{\rm eff}$, $\Delta N_{\rm eff}$, by its decay~\cite{DecayDR} is computed with $Y_\phi =n_\phi/s$, the actual number of particles per comoving volume and with the mass and life time of $\phi$ by using a simple relation~\cite{Scherrer:1987rr, Menestrina:2011mz}:
\begin{eqnarray}\label{DR_decay}
\Delta N^{\rm CMB}_{\phi~{\rm decay}} \simeq 8.3 (Y_\phi m_\phi/{\rm MeV})(\tau_\phi/{\rm s})^{1/2}.~~
\end{eqnarray}
Besides the dark radiation component coming from the $\phi$ decay, the primordial hidden photon $\gamma_H$ can contribute to $\Delta N_{\rm eff}$.
In this case, the primordial contribution is, however, quite suppressed as $\Delta N_{{\rm primo}~\gamma_H} \approx 0.053$, which is well explained in Ref.~\cite{Park:2013bza}.

\begin{figure}
\includegraphics[width=8.0cm]{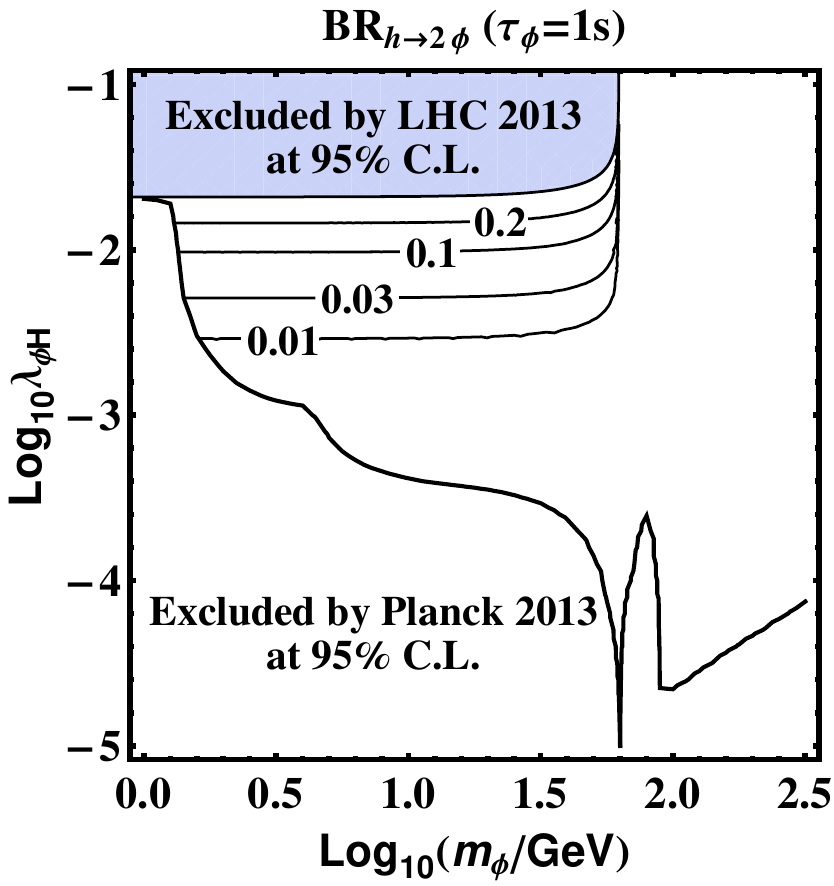}
\caption{Contour plots for ${\rm BR}_{h\to 2\phi}$ in the $m_\phi - \lambda_{\phi H}$ plane. 
The shaded region is constrained by the conservative invisible Higgs decay width limit ${\rm BR}_{\rm inv} < 0.34$ at the $95\%$ C.L.. 
The lower region below thick-solid line is excluded by the Planck limit on $N_{\rm eff}$ at the $95\%$ C.L. when $\tau_\phi = 1$s.
}
\label{fig2}
\end{figure}

Finally, the total contribution of this WIMPZILLA model to $N_{\rm eff}$ is given by
\begin{eqnarray}
\Delta N_{\rm eff}^{WZ} &=& N_{\rm eff}-N^{\rm SM}_{\rm eff}\nonumber \\
&=& \Delta N^{\rm CMB}_{\phi ~{\rm decay}} + \Delta N_{{\rm primo}~\gamma_H}\,,~~
\end{eqnarray}
which should be compared with the observational results in Eq.~\eqref{eq:neff}, in particular, the most stringent Planck 2013 result~\cite{Planck2013}. 
The lower region below thick-solid line in Fig.~\ref{fig2}, where $\tau_\phi = 1$ s is chosen as a reference value, is excluded by the Planck result at the $95\%$ confidence level (C.L.) due to the large contribution to $N_{\rm eff}$.
The dip around $m_\phi = m_h/2$ appears because of the resonant $s-$channel annihilation of $\phi$ into SM particles mediated by the Higgs boson.
$\Delta N^{\rm WZ}_{\rm eff}$ at the CMB and BBN epochs is discussed in detail in Ref.~\cite{Park:2013bza}.

\subsection{Invisible Decay of the Higgs}

In our scenario, we have another important possibility to test the model in collider experiments because the hidden scalar $\phi$ interacts with the SM Higgs boson $h$. 
The new decay channel of $h$ to $\phi$ is open due to the non-vanishing `Higgs-portal' interaction when kinematically allowed. 
The decay rate is given by
\begin{equation}
\Gamma_{h\to 2\phi} \simeq \frac{\lambda_{\phi H}^2}{32\pi} \frac{v^2}{m_h} \sqrt{1-\frac{4m_\phi^2}{m_h^2}}\,.
\end{equation}
Recently, the invisible branching fraction of the Higgs was measured by ATLAS~\cite{ATLAS} and CMS~\cite{CMS}.
A global fit analysis based on all the Higgs search data provides a constraint on the invisible branching fraction of the Higgs: 
${\rm BR}_{\rm inv}=\Gamma_{h\to {\rm invisible}}/\Gamma_{h\to {\rm all}} < 0.24$ at the $95\%$ C.L. assuming the SM decay rates for the other decay modes and ${\rm BR}_{\rm inv} < 0.34$ at the $95\%$ C.L. allowing non-standard values for $h \rightarrow \gamma\gamma$ and $h \leftrightarrow gg$~\cite{HiggsFit}.
We finally obtained a limit on the Higgs portal coupling which is shown in Fig.~\ref{fig2}.

\section{Conclusion}

WIMPZILLA with $m_{DM} \approx 10^{12-14}$ GeV can satisfy the required DM relic density when a new gauge symmetry U(1)$_H$ protects WIMPZILLA from decay as U(1)$_{em}$ for the electron. 
The new gauge boson of U(1)$_H$ can provide a possibility of testing this simple WIMPZILLA model by tracing dark radiation in the BBN and the CMBR data. 
In addition, we may (dis)prove  this model by measuring the invisible decay of the Higgs at collider experiments.

\begin{acknowledgments}
This work is supported by Basic Science Research Program through the
National Research Foundation of Korea funded by the Ministry of
Education, Science and Technology NRF-2013R1A1A2061561 (JC), NRF-2013R1A1A2064120 (SC).
We appreciate APCTP for its hospitality during completion of this work.
\end{acknowledgments}

\end{document}